\documentclass[a4paper]{jpconf}
\usepackage[utf8]{inputenc}

\usepackage{amsmath}
\usepackage{amssymb}
\usepackage{mathtools}
\usepackage{braket}
\usepackage{dsfont}
\usepackage{graphicx}
\usepackage{color}
\usepackage{cancel}
\usepackage{tensor}
\usepackage{stackrel}
\usepackage{cite}

\DeclareMathOperator{\diag}{diag}

\begin{document}

\title{Covariant bandlimitation from\\ Generalized Uncertainty Principles}

\author{J Pye}

\address{Department of Applied Mathematics, University of Waterloo, Waterloo, Ontario, N2L 3G1, Canada}
\address{Institute for Quantum Computing, University of Waterloo, Waterloo, Ontario, N2L 3G1, Canada}
\address{Perimeter Institute for Theoretical Physics, 31 Caroline St. N., Waterloo, Ontario, N2L 2Y5, Canada}

\ead{j2pye@uwaterloo.ca}

\begin{abstract}
  It is widely believed that combining the uncertainty principle with gravity will lead to an effective minimum length scale. A particular challenge is to specify this scale in a coordinate-independent manner so that covariance is not broken. Here we examine a class of Lorentz-covariant generalizations of the uncertainty principle which aim to provide an effective low-energy model for a Lorentz-invariant minimum length. We show how this modification leads to a covariant bandlimitation of quantum field theory. However, we argue that this does not yield an adequate regulator for many quantities of interest, e.g., the entanglement entropy between spatial regions. The possibility remains open that it could aid in regulating interactions.
\end{abstract}

\section{Introduction}

In 1936, Bronstein \cite{bronstein1936quantentheorie} suggested a fundamental limit to the precision of gravitational field measurements.
Ensuing thought experiments have indicated that in the presence of gravity, there should generally be a limitation to the precision of distance measurements in the form of a lower bound on the uncertainty, given by the Planck length: $\Delta X \geq \ell_P := \sqrt{\hbar G / c^3} \sim 10^{-35}m$ \cite{mead1964possible,hossenfelder2013minimal}.
It is widely believed this indicates that the classical conception of spacetime breaks down at the Planck scale.
As Bronstein \cite{bronstein1936quantentheorie,bronstein2012republication} writes: ``Without a deep revision of classical notions it seems hardly possible to extend the quantum theory of gravity also to this domain.''

Intuitively, the presence of a minimum length suggests that spacetime may be fundamentally discrete, rather than an infinitely-divisible continuum.
Naively, if one were to imagine this in analogy with a crystal lattice, one would seem to be forced to abandon the continuous spacetime symmetries fundamental to, for example, the structure of the Standard Model.
The spirit of the work presented in this paper is to examine whether one must abandon continuous symmetries by adopting a minimum length.
As it happens, this forfeiture is not generally required.

This endeavor is inspired by a model for a minimum length scale called the Generalized Uncertainty Principle (GUP) \cite{kempf1994uncertainty,kempf1995hilbert}, which directly introduces Planck-scale corrections to the ordinary quantum-mechanical uncertainty principle.
A distinguishing feature of this model is that it introduces a minimum length while preserving Euclidean symmetries (translation and rotation invariance).
Given this fact, it is natural to ask whether an extension of the GUP could be made to also include Lorentz symmetry.

Without a full working theory of quantum gravity, it is difficult to obtain a clear picture of the nature of this minimum length.
The goal here is to begin in the well-established territory of quantum field theory (QFT) and deform it toward the Planck scale while retaining Lorentz-invariance.
Our particular focus will be on developing Lorentz-covariant versions of the GUP and apply it to QFT.
The purpose is to clarify the role of the minimum length as a covariant regulator for QFT \cite{deser1957general,dewitt1964gravity}.
Of particular interest here is the entanglement entropy of the field, with the ultimate aim of studying black hole entropy.

In this paper, we will begin in Section~\ref{sec:gup} by reviewing the original version of the GUP.
In Section~\ref{sec:sampling}, we will show how the GUP relates to bandlimitation and sampling theory, which clearly demonstrates the possibility of discrete representations while retaining translation invariance.
In Section~\ref{sec:cov_gup}, we introduce Lorentz-covariant extensions of the GUP.
We then proceed to show in Section~\ref{sec:cov_bl} that one can arrive at a covariant analogue of bandlimitation, but it does not yield discrete spacetime representations as in the Euclidean case.
Lorentz-covariant bandlimitation has been considered before \cite{kempf2004covariant,kempf2013fully,chatwin2017natural}, in direct analogy with the Euclidean case.
The contribution we provide here is to show that one is lead to the same conclusion beginning with a covariant adaptation of the GUP.
In particular, we show that covariant bandlimitation is the only non-trivial possibility arising from the Lorentz-covariant GUP we study here.
We then provide further discussion of these results in Section~\ref{sec:discussion}.

Throughout we will work in Planck units, for which $\hbar = c = G = 1$.

\section{Generalized Uncertainty Principles}\label{sec:gup}

The idea behind the Generalized Uncertainty Principle (GUP) is to restrict the set of allowable states by directly modifying the ordinary uncertainty principle to impose a minimum position uncertainty, $\Delta X \geq \ell_P$.
(For purposes of simplicity, here we will discuss the one-dimensional version of the GUP, although it can be extended to higher dimensions \cite{kempf1994uncertainty,kempf1995hilbert}.)
The allowable states in ordinary quantum mechanics satisfy $\Delta X \Delta P \geq \tfrac12$.
A Planck-scale modification should become significant when the momentum uncertainty is large, since this will cause large uncertainties in the geometry through the gravitational field equations, which will in turn generate an increase in position uncertainty.
The GUP achieves this by modifying the uncertainty principle to contain gravitational corrections, for example, to lowest order of the form \cite{kempf1994uncertainty,kempf1995hilbert}:
\begin{equation}
  \Delta X \Delta P \geq \tfrac12 ( 1 + \ell_P^2 \Delta P^2 ).
\end{equation}
The physical states must satisfy this modified uncertainty principle.
It is easy to show that the above inequality implies $\Delta X \geq \ell_P$.
There are many other modifications which yield a finite minimum uncertainty, but generically each exhibits the same qualitative behavior \cite{kempf1997non}.

One can arrive at a GUP of the above form through a modification of the commutation relation between the operators $X$ and $P$ \cite{kempf1994uncertainty,kempf1995hilbert}:
\begin{equation}
  [ X, P ] = i ( 1 + \ell_P^2 P^2 ).
\end{equation}
This implies an uncertainty principle of the above form (up to a term proportional to $\langle P \rangle^2$, although this does not significantly change the above conclusion).
Hence, if we consider states in a representation of this modified Heisenberg algebra, they will automatically exhibit the feature of a minimum uncertainty in position.
For example, one such representation is:
\begin{equation}
  (X \tilde{\psi})(p) = i ( 1 + \ell_P^2 p^2 ) \frac{d}{dp} \tilde{\psi}(p), \qquad (P \tilde{\psi})(p) = p \tilde{\psi}(p),
\end{equation}
with inner product, $\langle \tilde{\phi} | \tilde{\psi} \rangle = \int \frac{dp}{2\pi} (1 + \ell_P^2 p^2)^{-1} \tilde{\phi}^\ast(p) \tilde{\psi}(p)$.
The physical states are those in a dense subset of this Hilbert space (typically the common domain of $[X,P]$ and various powers of $X$, $P$, and their products).

There have been many papers studying various implications of GUPs (e.g., see \cite{kempf1997non,brau1999minimal,hossenfelder2003signatures,das2008universality,pikovski2012probing,scardigli2015gravitational,scardigli2018modified}), including extensions to relativistic quantum mechanics (with a non-commutative geometry) \cite{todorinov2019relativistic}.
Our interest in this paper is to study Lorentz-covariant modifications of QFT as one approaches the Planck scale.
In order to apply the GUP to QFT, one must first identify how the ordinary uncertainty principle of non-relativistic quantum mechanics manifests itself in QFT.
After making this identification and performing the appropriate generalization to the GUP, we will show how such a modification can be interpreted in terms of bandlimitation and sampling theory.

\section{Bandlimitation and sampling theory}\label{sec:sampling}

The classical space of field configurations (on flat spacetime) is taken to be a representation space of the Poincar\'e group.
From the Poincar\'e generators $L_{\mu \nu}$ and $P^\lambda$, one can build momentum space representations of the fields, $\tilde{\phi}(p) \equiv ( p | \tilde{\phi} )$, with $P^\lambda | p ) = p^\lambda | p )$.
(Note that we use round brackets to emphasize the distinction from quantum mechanical wavefunctions.)
The position space or spacetime representation of the fields is then naturally obtained via a Fourier transform.
The assumption that position and momentum space are Fourier-related presumes an underlying Heisenberg algebra structure, $[ X_\mu, P^\nu ] = i \tensor{\delta}{_\mu^\nu}$, with position representation $\phi(x) \equiv (x|\phi)$, where $X^\mu |x) = x^\mu |x)$.
This is formally similar to the typical account in quantum mechanics.
Naturally, this is the step at which one can introduce a modification of the Heisenberg algebra.
That is, one can continue to build momentum space representations for the Poincar\'e group, but a modified Heisenberg algebra may alter the relationship between position and momentum space.

Notice that in the position representation, $x$ is simply a label for an element in the joint spectrum of the $X^\mu$ operators.
Hence, field configurations in the spacetime representation reside on the joint spectra of these operators.
A modification of the Heisenberg algebra may change the structure of these spectra.
For example, if the spectra become discrete after a modification, then these field configurations could be thought of as living on a lattice.

Now we will consider more explicitly such a modification using the GUP discussed in Section~\ref{sec:gup}.
For simplicity, we will continue with the one-dimensional case at a fixed time (however, the Lorentz-covariant GUP in Section~\ref{sec:cov_gup} will apply to arbitrary spacetime dimensions).
In the previous section, we presented the momentum space representation of the GUP corresponding to $[ X, P ] = i ( 1 + \ell_P^2 P^2 )$.
In order to find the position space representation, we must find the eigenvalues and eigenvectors of $X$.
However, the finite minimum uncertainty, $\Delta X \geq \ell_P$, implies that $X$ has no eigenvectors in the domain where the GUP is represented, since the uncertainty for an eigenvector would vanish (or in the case of a continuous spectrum could be made arbitrarily close to zero).

This situation can be made more clear by performing a momentum space diffeomorphism.
Using the functional calculus of $P$, we define an operator, $K := \frac{1}{\ell_P} \arctan ( \ell_P P )$, which can be easily shown satisfies the canonical commutation relation with $X$: $[ X, K ] = i$.
It may appear that this simply reverts the GUP to the ordinary uncertainty principle, with $x$-space and $k$-space being Fourier-related.
However, because $\arctan$ has a finite range, the fields in $k$-space have finite support (such fields are called \emph{bandlimited}).
The $k$-space representation consists of fields on the interval $[-\frac{\pi}{2\ell_P},\frac{\pi}{2\ell_P}]$ obeying Dirichlet boundary conditions (and with the usual flat measure $dk/2\pi$).
Therefore, by focusing solely on the $x$- and $k$-space representations, we can view the deformation as equivalent to simply restricting to a subset of $k$-space.
In other words, we are restricting the spectrum of the $K$ operator (ordinarily represented in position space as the derivative operator $-i d/dx$).
Note that one can also perform this diffeomorphism when considering the GUP for wavefunctions in quantum mechanics, and think of the resulting system as a particle in a box in momentum space.

Do we still have finite minimum uncertainty in position?
Does the position operator continue to lack eigenvectors?
Of course this should be the case, since we have not changed the abstract algebraic properties enforcing these features.
Although the operators $X$ and $K$ are canonically related, the bandlimitation implies a maximum uncertainty in $K$.
Therefore, the uncertainty in $X$ continues to exhibit a finite minimum since $\Delta X \geq 1/2\Delta K \geq 1/2\Delta K_{\text{max}}$.
The situation involving the eigenvectors is more subtle, for it turns out that $X$ is symmetric on the physical domain, but not essentially self-adjoint \cite{kempf1994uncertainty,kempf1995hilbert}.
Although the formal eigenvectors of $X$ in $k$-space are $e^{-i k x}$, these do not obey the appropriate Dirichlet boundary conditions.
In order to obtain a position space representation, one needs to extend the domain of physical states to a domain on which $X$ is self-adjoint.
However, in this case the choice of extension is not unique; indeed, there is a one-parameter family of self-adjoint extensions.
These correspond to the periodic boundary conditions up to a phase $e^{-2\pi i \alpha}$, where $\alpha \in [0,1)$ parametrizes the family of extensions.
For each extension $\alpha$, one obtains a discrete spectrum $\{ x_n^{(\alpha)} = 2 \ell_P (n+\alpha) \}_{n \in \mathds{Z}}$ with corresponding eigenvectors.
Because the physical domain is contained within the domain of each of these extensions, any one of the set of eigenvectors can be used as a basis in which one can represent the physical fields: $|\phi) = \sum_{n \in \mathds{Z}} |x_n^{(\alpha)})(x_n^{(\alpha)}|\phi)$.
Therefore, these fields admit a family of discrete representations in position space.
Furthermore, although we have discrete representations, we also retain full translation invariance in space because none of the lattices are preferred.

This situation can be illustrated in a much simpler scenario.
If we introduce an ultraviolet cutoff for the space of $L_2$ functions on a circle, then we obtain a finite-dimensional function space (finitely many Fourier modes).
Suppose we then pick a set of sample points on the circle, equal in number to the number of Fourier modes.
Generically, we can use the function values at these points to completely specify a function in the space.
We can then view this as a representation of the functions on this lattice of sample points, although we have not changed the fact that the underlying space remains continuous.

The discreteness of the position representations of the bandlimited fields is analogous to this example, except extended to the real line and an infinite-dimensional function space.
Notice that the sample points $x_n^{(\alpha)} = 2\ell_P(n+\alpha)$ cover the real line exactly once as one runs through both $n \in \mathds{Z}$ and $\alpha \in [0,1)$.
Hence, we can consider a field value at an arbitrary $x \in \mathds{R}$ as $\phi(x) \equiv ( x | \phi )$, since $|x) = |x_n^{(\alpha)})$ for some $n$ and $\alpha$.
Therefore, we can consider position space to be the entire real line, with the fields admitting discrete representations on these sampling lattices.
The field value at any other point on the real line can be obtained via the following reconstruction formula:
\begin{equation}
  \phi(x) = (x|\phi) = \sum_{n \in \mathds{Z}} (x|x_n^{(\alpha)})(x_n^{(\alpha)}|\phi) = \sum_{n \in \mathds{Z}} \frac{ \sin [ \pi (x-x_n^{(\alpha)}) / 2 \ell_P ] }{ \pi (x-x_n^{(\alpha)}) / 2 \ell_P } \phi(x_n^{(\alpha)}),
\end{equation}
which is the basic result of Shannon sampling theory \cite{shannon1948mathematical,shannon1949communication,nyquist1928certain}.
The quantization of these bandlimited fields is relatively straightforward \cite{kempf1994quantum,kempf1997quantum,kempf2000fields,pye2015locality}.
Also, many of these results have been extended to higher-dimensional spaces \cite{landau1967necessary,jerri1977shannon} as well as curved Riemannian manifolds \cite{kempf2004covariant,kempf2008information}.

The main message of sampling theory is that one can have discrete representations while retaining continuous Euclidean spatial symmetries.
Armed with this encouragement, we will now examine whether this can be extended to Lorentzian spacetime symmetries.

\section{Lorentz-covariant Generalized Uncertainty Principles}\label{sec:cov_gup}

The strategy of the GUP was to introduce a minimum length scale by modifying the commutation relation between $X$ and $P$ to give a finite minimum uncertainty in position.
Now we will attempt to do this Lorentz-covariantly.
By \emph{Lorentz-covariant}, we mean that the coordinates, and not necessarily the momenta, transform appropriately under an infinitesimal Lorentz transformation.
This is because we are primarily concerned with symmetries of spacetime, rather than the space described by the momenta.
Although there are known problems with deformations of momentum space (e.g., the so-called ``soccer-ball problem'' \cite{hossenfelder2014soccer,hossenfelder2013minimal}), we will not address these issues here.

We will continue to focus primarily on the kinematical structure of the fields (although dynamics will be briefly discussed in Section~\ref{sec:discussion}).
Indeed, if we are interested in a regulator for the entanglement entropy, it should arise from the kinematics, as entanglement is typically thought of as between regions of space at a fixed time.

We will consider a somewhat general set of deformations of the Heisenberg algebra:
\begin{equation}
  [ X_I, P^\mu ] = i \tensor{\theta}{_I^\mu}(P), \qquad [ X_I, X_J ] = 0, \qquad [ P^\mu, P^\nu ] = 0.
\end{equation}
where we assume $\tensor{\theta}{_I^\mu}$ is a smooth function of the $P^\mu$ operators, and as a matrix is pointwise-invertible (i.e., $\det \theta \neq 0$).
(Note that $\tensor{\theta}{_I^\mu}$ should be a function of the dimensionless quantity $\ell_P P^\mu$, however, to keep the notation compact, we will omit an explicit reference to $\ell_P$.)
We take $\tensor{\theta}{_I^\mu}$ to be a function of the $P^\mu$ operators and not the $X_I$ operators for both simplicity and in analogy with the one-dimensional GUP (where $X$-dependence yields an infrared, rather than an ultraviolet, cutoff).
Both sets of indices traverse all spacetime dimensions, $I, \mu = 0,1,2,\dots,n-1$.
We label the $X_I$ operators and $P^\mu$ operators differently since we only require the upper-case Latin indices to transform appropriately under Lorentz transformations.

The requirements of commuting coordinates and commuting momenta are made solely as simplifying assumptions in order to make conclusive statements.
If this prevents the model from producing a suitable cutoff, then one could take this as an indication that these assumptions (such as the commutative geometry assumption) should be relaxed in order to obtain a cutoff in a model of this kind.
Indeed, we are attempting to determine whether one can obtain a minimum length without resorting to non-commutative geometry.

Of course, we will also assume that the Jacobi identities are satisfied.
This introduces the following restriction for $\theta$: $[ X_I, \tensor{\theta}{_J^\mu}(P) ] = [ X_J, \tensor{\theta}{_I^\mu}(P) ]$.
With this structure, it is simple to build a $p$-space representation for this algebra as:
\begin{equation}
  (X_I \phi)(p) = i \tensor{\theta}{_I^\mu}(p) \frac{\partial}{\partial p^\mu} \phi(p), \qquad (P^\mu \phi)(p) = p^\mu \phi(p),
\end{equation}
with inner product: $(\phi | \psi) = \int \frac{d^np}{(2\pi)^n} (\det \theta(p))^{-1} \phi^\ast(p) \psi(p)$.
Note that throughout we assume the summation convention.
In this representation, the condition for the Jacobi identity becomes: $\tensor{\theta}{_I^\nu} \partial_\nu \tensor{\theta}{_J^\mu} = \tensor{\theta}{_J^\nu} \partial_\nu \tensor{\theta}{_I^\mu}$.
This ensures that the coordinates commute and are symmetric operators.
One can write the Lorentz generators in this representation as:
\begin{equation}
  L_{IJ} := X_I \int^p dp'^\mu \tensor{(\theta^{-1})}{_\mu_J}(p') - X_J \int^p dp'^\mu \tensor{(\theta^{-1})}{_\mu_I}(p'),
\end{equation}
where the upper-case Latin indices are lowered using the Minkowski metric, $\eta_{IJ} := \diag(+1,-1,-1,\dots,-1)$.
It is straightforward to show that these indeed satisfy the Lorentz algebra, and that the $X_I$ operators transform appropriately as: $[ X_K, L_{IJ} ] = i \eta_{JK} X_I - i \eta_{IK} X_J$.

In the one-dimensional case, we made a simplification by enacting a coordinate change in momentum space to recover canonical commutation relations, but with a restriction of the spectrum of the new momentum operator.
We can also establish such a diffeomorphism here for an arbitrary deformation $\tensor{\theta}{_I^\mu}(P)$ (with the already-mentioned assumptions).
Notice that the operators $-i X_I$ provide a basis of vector fields on $p$-space.
Because these vector fields commute with each other, they can be written as a coordinate basis after a suitable diffeomorphism (see, e.g., \cite{lee2002introduction}).
That is, there is some diffeomorphism $k^I(p)$ for which we can write $X_I \phi(k) = i (\partial / \partial k^I) \phi(k)$.
The functions parametrizing the deformation of the Heisenberg algebra can then simply be associated with the (inverse) Jacobian of this diffeomorphism, $\tensor{\theta}{_I^\mu}(p) = (\partial k^I(p) / \partial p^\mu)^{-1}$.
We can define multiplication operators in $k$-space as $K^I \phi(k) = k^I \phi(k)$ (or in $p$-space as $K^I \phi(p) = k^I(p) \phi(p)$), which satisfy: $[ X_I, K^J ] = i \tensor{\delta}{_I^J}$.
Notice that with the $K^I$ operators, the Lorentz generators take their familiar form: $L_{IJ} = X_I K_J - X_J K_I$.

\section{Covariant bandlimitation}\label{sec:cov_bl}

We have established that a deformation of the commutator between $X_I$ and $P^\mu$ in the Heisenberg algebra can be reverted to the canonical commutation relations (provided the coordinate and momentum operators remain commuting among themselves).
Therefore, in regards to the kinematical structure of the classical field configurations, we are only left with the possibility of modifying the global properties of momentum space.
Can we restrict to a subset of $k$-space in order to achieve a kind of bandlimitation, as in the one-dimensional case?

Of course, we cannot make an arbitrary restriction.
A faithful representation of the Lorentz group in $k$-space requires the hypersurfaces $k^2 = \text{const}$ to remain untouched.
This leaves only the possibility of restricting $k$-space to a subset of these hypersurfaces, i.e., we can restrict the set of admissible mass shells.
For example, we can constrain the quantity $|k^2| = |k_0^2 - \boldsymbol{k}^2| < 1/\ell_P^2$.
This can be achieved, for example, by the following deformation:
\begin{equation}
  \tensor{\theta}{_I^J}(p) = f(p^2) \tensor{\delta}{_I^J} + p_I p^J \frac{2 f(p^2) f'(p^2)}{f(p^2) - 2 p^2 f'(p^2)},
\end{equation}
where $p^2 := p_0^2 - \boldsymbol{p}^2$, with diffeomorphism $k^I = p^I/f(p^2)$ and inverse denoted $p^I = b(k^2) k^I$.
To achieve the desired restriction of $k^2$, we can choose $b(k^2) = \sum_{n=0}^\infty (\ell_P^2 k^2)^{2n}$.
Due to the finite radius of convergence of this series, fields in the domain of $P^I$ are confined to the region $|k^2| < 1/\ell_P^2$.
One can check that this indeed defines a diffeomorphism in this region, and by writing $\tensor{\theta}{_I^J}(p(k)) = b(k^2) \tensor{\delta}{_I^J} + 2 k_I k^J b'(k^2)$, that $\theta$ satisfies the required properties.

In the literature, such a restriction of momentum space has been called ``covariant bandlimitation'' \cite{kempf2004covariant,kempf2013fully,chatwin2017natural}.
In these previous works, this was considered as a restriction of the spectrum of the covariant d'Alembertian operator, in analogy with the restriction of the spectrum of $-i d/dx$ in the one-dimensional case.
Here we have shown that one is also led to covariant bandlimitation by a class of deformations of the Heisenberg algebra.

Despite the analogy with the Euclidean case, covariant bandlimitation does not exhibit the discrete spacetime representations characteristic of Shannon sampling theory.
The fact that this cannot occur is a consequence of a theorem of Landau \cite{landau1967necessary}, which states that the required density of sampling points is bounded below by the volume of the support of the space of functions in momentum space.
If we require a faithful representation of the Lorentz group in this space, this volume must be non-compact.
This immediately implies an infinite density of degrees of freedom.
(It is possible to derive a type of reconstruction formula for covariantly bandlimited fields, although the features are qualitatively different than the Euclidean case \cite{kempf2004covariant,kempf2013fully}.)

\section{Discussion and conclusion}\label{sec:discussion}

Despite the infinite density of degrees of freedom, what are the implications of a covariant bandlimit?
Does it provide a regulator for field theory?
It is clear that any on-shell quantities, such as the Wightman function, commutator, and correlation functions, are unaffected by the covariant bandlimit (for masses smaller than the cutoff) since these mass shells remain unchanged.
Hence, for a free field theory, it has limited impact as many physical quantities can be calculated on-shell.
For example, the entanglement entropy in the ground state between spatial regions can be calculated from the commutator and correlation functions (see, e.g., \cite{sorkin1983entropy,bombelli1986quantum,serafini2017quantum}).

Of course, the notion of ``on-shell'' presumes a form for the dynamics, such as that associated with the Klein-Gordon action, $S[\phi] = \tfrac12 ( \phi | K_I K^I \phi ) - \tfrac12 m^2 ( \phi | \phi )$.
However, one could also specify an action in terms of the additional set of momentum operators, $P^\mu$.
For example, a natural generalization could be $S[\phi] = ( \phi | \sigma(P,p_0) \phi ) - ( \phi | \sigma(p_m,p_0) \phi )$, where $\sigma$ is the Synge world function in $p$-space, and we define $p_0$ and $p_m$ as points in $p$-space which satisfy $k^I(p_0) = 0$ and  $k_I(p_m) k^I(p_m) = m^2$.
Due to the coordinate-invariance of $\sigma$, if we choose the $p$-space metric to be the pullback of the Minkowski metric on $k$-space, that is, $g_{\mu \nu}(p) = \eta_{IJ} \tensor{(\theta^{-1})}{_\mu^I} \tensor{(\theta^{-1})}{_\nu^J}$, then this simply reduces to the Klein-Gordon action.
One could alternatively introduce another metric on $p$-space for this purpose.
For example, with the deformation used in the previous section (with $p^I = b(k^2) k^I$), one could use $\eta_{IJ}$ in $p$-space to obtain the action: $S[\phi] = \tfrac12 ( \phi | P_I P^I \phi ) - \tfrac12 b(m^2)^2 m^2 ( \phi | \phi )$.
This can be viewed as a replacement of the d'Alembertian operator $\Box \to \Box/(1 - \ell_P^4 \Box^2)^2$ along with a re-scaling of the mass.
However, regardless of the choice, to retain covariance in $x$-space one is restricted to using functions of the d'Alembertian operator in the action.
A locally analytic function whose power series has a finite radius of convergence will yield covariant bandlimitation \cite{kempf2004covariant}.
Viewed in this way, the fact that covariant bandlimitation does not regulate the entanglement entropy is consistent with previous investigations in non-local field theories and modified dispersion relations \cite{nesterov2011gravitational,solodukhin2011entanglement}.
(Also, certain deformations considered here could be viewed as equivalent to non-local d'Alembertian operators arising from averaging over causal sets \cite{aslanbeigi2014generalized,belenchia2015nonlocal}.)

If we focus solely on the effect of covariant bandlimitation (and retain the usual Klein-Gordon action), then on-shell quantities are not modified, but the covariant bandlimit can be interpreted as cutting off allowable virtual masses.
Therefore, this will modify the inhomogeneous Green's functions of the theory, such as the Feynman propagator \cite{kempf2013fully,chatwin2017natural}.
Covariant bandlimitation may then be of interest in interacting theories.
For example, a simple interacting theory one could consider is that between the field and an accelerating detector, in which one can study the Unruh effect \cite{unruh1976notes}.
However, because the Unruh effect corresponds to a first-order response, there are no internal field lines which could be affected by the covariant bandlimit.
Therefore, there is no modification to the usual leading-order thermal response of an accelerated detector.
One could also consider interacting fields, although bandlimited functions do not form an algebra with pointwise multiplication in spacetime.
Thus, to retain a consistent mathematical description of interacting bandlimited fields, one should alter the multiplication by modifying the vertex.

The conclusions regarding covariant bandlimitation also hold on other spacetimes.
On a general pseudo-Riemannian manifold, covariant bandlimitation can be analogously defined in terms of cutting off the spectrum of a covariant differential operator, such as the Laplace-Beltrami operator, $\Box_g = |g|^{-1/2} \partial_\mu ( |g|^{1/2} g^{\mu \nu} \partial_\nu \cdot )$ \cite{kempf2004covariant,kempf2013fully,chatwin2017natural}.
However, even in this case, the eigenvalues correspond to allowable masses in the Klein-Gordon equation, $\Box_g \phi = - m^2 \phi$.
Therefore, restricting the allowable masses has no effect on the homogeneous solutions to the Klein-Gordon equation, and thus cannot be used to regulate the entanglement entropy (for example, between the interior and exterior regions of a black hole).
This last observation is also consistent with the investigations of \cite{nesterov2011gravitational,solodukhin2011entanglement}.

We noted above that the obstruction to obtaining discrete spacetime representations was due to the infinite volume of momentum space.
If we consider restricting to a finite volume of momentum space (through whatever means), then in spacetime one obtains a space of analytic functions (see, e.g., \cite{rudin1987real}).
Analytic functions exhibit a severe non-locality since all of the degrees of freedom are contained in a small volume around any point in spacetime (as the Taylor series can be evaluated in a neighborhood).
(For a related discussion, see also \cite{pye2015locality}.)
Hence it seems that anything akin to ordinary sampling theory would allow for superluminal signalling, thus would be inappropriate for the configuration space of a relativistic field theory.
In practical applications of sampling theory (e.g., in communications engineering \cite{shannon1948mathematical,shannon1949communication,nyquist1928certain}) this non-locality does not appear as realistic signals are only approximately bandlimited.
In any case, practically one could not fully evaluate the Taylor series coefficients to arbitrary precision.
Therefore, instead of attempting to model the fundamental configuration space of the field theory, one could adopt these analytic functions as an approximation, and perhaps prevent superluminal effects in the model by introducing other practical limitations (such as finite precision of measurements).

To conclude, although the original GUP yields a sampling theory for bandlimited fields, we have shown that in a class of Lorentz-covariant GUPs one does not obtain a finite density of degrees of freedom in spacetime.
Consequently, quantities such as entanglement between spatial subsystems is not regulated, although covariant bandlimitation may still be pertinent in interacting theories.
In order to obtain a stronger regulator, perhaps certain assumptions should be relaxed, for example, the commutativity of the coordinates (see, e.g., the reviews \cite{douglas2001noncommutative,szabo2003quantum}).

\ack{
The author would like to thank Achim Kempf for his supervision, continuing support, and many invaluable discussions.
The author also acknowledges support from the Natural Sciences and Engineering Research Council (NSERC) of Canada through the Doctoral Canada Graduate Scholarship (CGS) program, as well as from the Ontario Graduate Scholarship (OGS) program.
This research was supported in part by Perimeter Institute for Theoretical Physics. Research at Perimeter Institute is supported by the Government of Canada through the Department of Innovation, Science and Economic Development and by the Province of Ontario through the Ministry of Research and Innovation.
}

\section*{References}
\bibliographystyle{iopart-num}
\bibliography{main}

\end{document}